\documentclass[
	a4paper, %
	10pt, %
	unnumberedsections, %
	twoside, %
]{LTJournalArticle}

\usepackage{xcolor}
\usepackage[detect-all, per-mode=symbol, range-units=single, range-phrase=~to~]{siunitx}
\usepackage{glossaries}
\usepackage{cleveref}
\usepackage{subcaption}
\usepackage{url}

\titlespacing*{\section}{0pt}{3.1mm}{1.9mm}
\newacronym{sd}{SD}{Secure Digital}
\newacronym{sdhc}{SDHC}{SD host controller}
\newacronym{sdhci}{SDHCI}{SD host controller interface}
\newacronym{oseda}{OSEDA}{open-source electronic design automation}
\newacronym[longplural={systems on chip}]{soc}{SoC}{system on chip}
\newacronym{sbc}{SBC}{single-board computer}
\newacronym{cmo}{CMO}{cache management operation}
\newacronym{dma}{DMA}{direct memory access}
\newacronym{pdk}{PDK}{process design kit}
\newacronym{os}{OS}{open-source}

\newcommand{\lb}[1]{\textcolor{black}{#1}}

\newcommand{\riscv}{\mbox{RISC-V}}

\runninghead{} %

\footertext{\textit{RISC-V Summit Europe, Bologna, 8-12th June 2026}} %

\title{Implementing and Optimizing an Open-Source\\ SD-card Host Controller for RISC-V SoCs}

\author{%
	Axel Vanoni\textsuperscript{1}%
    \thanks{Corresponding author: \href{mailto:axvanoni@ethz.ch}{\tt axvanoni@ethz.ch}}, %
    Philippe Sauter\textsuperscript{1}, %
    Paul Scheffler\textsuperscript{1}, %
    Anton Buchner\textsuperscript{1}, \\%
    Micha Wehrli\textsuperscript{1}, %
    Thomas Benz\textsuperscript{1,2}, %
    and Luca Benini\textsuperscript{1,3} %
}

\date{\footnotesize\textsuperscript{\textbf{1}}~Integrated Systems Laboratory, ETH Zurich\\
\textsuperscript{\textbf{2}}~lowRISC C.I.C., Cambridge, United Kingdom\\
\textsuperscript{\textbf{3}}~Department of Electrical, Electronic, and Information Engineering, University of Bologna}

\renewcommand{\maketitlehookd}{%
\begin{abstract}
\noindent %
Recent \lb{announcements have shown} the viability of end-to-end \gls{os} Linux-capable {\riscv} \glspl{soc}. 
However, practical application and software development platforms require efficient non‑volatile storage, which is \lb{not adequately} served by common SPI-based interfaces due to their limited throughput.
\Gls{sd} cards are the de facto standard storage medium for embedded Linux systems; efficient \gls{sdhc} integration is thus essential for \lb{\glsdesc{os}} {\riscv} platforms.
We present an \gls{os} \gls{sdhci} peripheral integrated into the \lb{end-to-end \gls{os}} Cheshire {\riscv} \gls{soc} platform.
The controller and its software stack are designed \lb{with full awareness} of CVA6's memory system and Linux driver behavior; during evaluation, we identify a significant performance bottleneck caused by the {\riscv} memory model and CVA6's implementation of the fence instruction, which flushes the pipeline and data cache on memory-mapped register accesses when \glspl{cmo} are unavailable.
By customizing the driver’s register access paths and avoiding unnecessary fences, we substantially \lb{reduced} this overhead.
\lb{Our fully \gls{os}} controller achieves up to 11.1 MB/s throughput, approaching the 12.5 MB/s limit of the \gls{sd} interface and providing up to 6.5 times the throughput of SPI-based storage.
\end{abstract}
}

\begin{document}
\maketitle %

\glsresetall

\section{Introduction}

Recently, \gls{os} silicon efforts have demonstrated that multi-million-gate, Linux-capable {\riscv} \glspl{soc} can be realized using \gls{oseda} tools such as Yosys and OpenROAD, implemented using open \glspl{pdk}~\cite{sauter2025basilisk}.
Basilisk, a \SI{34}{\milli\metre\squared}, \SI{2.7}{\mega{GE}} silicon demonstrator in IHP's \SI{130}{\nano\metre} open \gls{pdk} proves the feasibility of end-to-end \gls{os} Linux-booting {\riscv} \glspl{sbc}~\cite{sauter2025basilisk}.
Building on the silicon-proven Cheshire \gls{soc} platform~\cite{ottaviano_cheshire_2023}, Basilisk features a CVA6 core~\cite{zaruba_cost_2019}, a dual-chip HyperBus memory interface, \SI{64}{\kilo\byte} SRAM, and a rich set of peripherals including SPI, I2C, UART, VGA, and a USB 1.1 host controller.
Yet, a key ingredient of a practical \gls{sbc} remains insufficiently addressed: Basilisk's support for persistent, high-capacity storage is limited by SPI or I2C throughput.
While {HyperRAM} enables Linux execution, non-volatile block storage is essential for booting, root file systems, and keeping user data\lb{. In} most embedded Linux \glspl{sbc}, \gls{sd} cards provide this functionality due to their low cost and commodity interface.

To address this gap, we implemented and integrated an \gls{sdhc} into Cheshire, enabling the next generation of Cheshire-based \glspl{sbc} to access persistent storage with higher throughput and efficiency.
\lb{Our contributions are}: \\
\vspace{-3mm}
\begin{itemize}
    \item An optimized implementation of an \gls{sdhc}, \gls{sdhci} version 1.0, and integration into the Cheshire \gls{soc} platform.
    \item An optimized \gls{sdhci} driver tailored for the CVA6 configuration currently used in Cheshire.
    \item Area and performance comparisons of our \gls{sdhc} implementation \lb{versus} \gls{sd} access over SPI.
    \item \lb{The synthesizable RTL code of our \gls{sdhc} is available under a permissive open-source license.~\footnote{\url{https://github.com/Freakness109/sdhci}}}
\end{itemize}

\section{Architecture and Integration}

\begin{figure}[t]
    \centering
    \begin{subcaptionblock}{\linewidth}
        \centering
        \includegraphics[width=\linewidth, height=2cm]{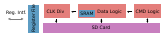}
        \vspace{-7mm}
        \caption{}
        \vspace{2mm}
        \label{fig:arch:sdhc}
    \end{subcaptionblock}
    \hfill
    \begin{subcaptionblock}{\linewidth}
        \centering
        \includegraphics[width=\linewidth]{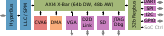}
        \vspace{-6mm}
        \caption{}
        \label{fig:arch:integration}
    \end{subcaptionblock}
    \vspace{-6mm}
    \caption{\Gls{sdhc} architecture (a) and its implementation in Cheshire (b).}
    \label{fig:arch}
\end{figure}

The architecture of the \gls{sdhc} is directly governed by the \gls{sd} bus protocol:
Since the command and data channels are independent, we implement them as separate architectural blocks, as shown in \Cref{fig:arch:sdhc}.
We derive the \gls{sd} clock from the system clock with a programmable integer clock divider.
The command module sends command-completion notifications to the data module to synchronize the two channels as needed.
Further, the data module can stop the \gls{sd} clock if its internal SRAM buffers fill faster than the host can empty them during reads.
To program the command, data, and clock blocks, we use the industry-standard approach of memory-mapped registers and implement the \gls{sdhci} standard, which specifies the register map and interrupt timings.

Most register-interface-based peripherals in Cheshire are connected to a separate register bus multiplexer, as shown in \Cref{fig:arch:integration}.
Not so the \gls{sdhc}: due to sensitivity to latency in data transfers, we connect the peripheral directly to the main AXI crossbar.
The latency sensitivity stems from CVA6, in its current parametrization in Cheshire, not supporting multiple outstanding requests to memory-mapped registers.
We reduced read access latency to eleven cycles, enabling a 4-byte transfer every 29 cycles in a bare-metal copy loop.
Writes are less taxed by latency, allowing one write every nine cycles.

\section{Driver Support}

Support for \gls{sdhci} facilitates driver development:
On Linux, we leverage the existing \gls{sdhci} driver to enable support for our peripheral in fewer than \SI{100}{LoC} of additional code.
During Linux driver verification, we noticed significant slowdowns compared to bare-metal execution.
The cause is CVA6's \emph{fence} instruction when \glspl{cmo} are not enabled.
Fences currently flush the processor pipeline and the data cache, introducing around 500 cycles of overhead for every 32-bit access to a memory-mapped register space.
To mitigate this, we override pre-existing driver callbacks that allow customizing register accesses to skip fences, tailoring the driver to CVA6.

Additionally, we implemented a bare-metal driver for booting from \gls{sd} cards, reducing the contribution to the Cheshire boot ROM compared to the current SPI driver implementation.

\section{Results}

\begin{figure}
    \centering
    \begin{subcaptionblock}{\linewidth}
        \begin{subcaptionblock}{0.45\linewidth}
            \centering
            \includegraphics[width=\linewidth]{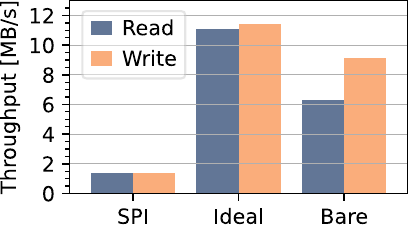}
            \vspace{-5.5mm}
            \caption{Bare Metal}
            \vspace{2mm}
            \label{fig:throughput:bare}
        \end{subcaptionblock}
        \hspace{1mm}
        \begin{subcaptionblock}{0.45\linewidth}
            \centering
            \includegraphics[width=\linewidth]{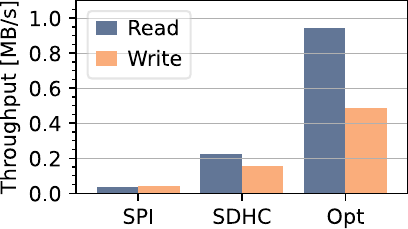}
            \vspace{-5.5mm}
            \caption{\SI{50}{\mega\hertz}}
            \vspace{2mm}
            \label{fig:throughput:linux}
        \end{subcaptionblock}
    \end{subcaptionblock}
    \begin{subcaptionblock}{\linewidth}
        \begin{subcaptionblock}{0.45\linewidth}
            \centering
            \includegraphics[width=\linewidth]{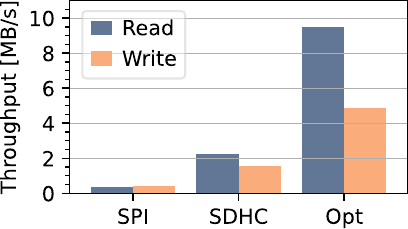}
            \caption{Scaled (\SI{500}{\mega\hertz})}
            \label{fig:throughput:linux-scaled}
        \end{subcaptionblock}
        \hspace{1mm}
        \begin{subcaptionblock}{0.45\linewidth}
            \centering
            \includegraphics[width=\linewidth]{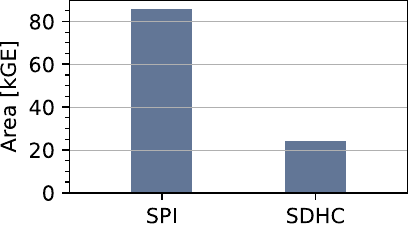}
            \caption{Peripheral Area}
            \label{fig:area}
        \end{subcaptionblock}
    \end{subcaptionblock}
    \vspace{-4.75mm}
    \caption{\Gls{sdhc} throughput on bare-metal (a) and on Linux (b-c); Area results (d).}
    \label{fig:results}
\end{figure}

We present two sets of performance benchmarks shown in \Cref{fig:results}:
First, we transfer 16 blocks, \SI{512}{\byte} each, with the host system running at \SI{50}{\mega\hertz} and the \gls{sd} card running at \SI{25}{\mega\hertz}.
We run transfers with the stand-alone peripheral with single-cycle register access (\emph{Ideal}) and with the peripheral integrated into Cheshire (\emph{Bare}).
Second, we present performance under Linux before (\emph{SDHC}) and after optimizations (\emph{Opt}), at the same operating frequencies as above: %
We transfer \SI{4}{\mega\byte} using \emph{dd} to read from and write to a mounted ext4 partition (\emph{fsync} and bypassing disk cache).
We extrapolate the performance to a Cheshire tape-out scenario with the host running at \SI{500}{\mega\hertz} and the \gls{sd} card at \SI{25}{\mega\hertz}.
In both benchmarks, we show the SPI peripheral as a baseline.

In the \emph{Ideal} scenario, the \gls{sdhc} has a throughput of \SI{11.1}{\mega\byte\per\second} in reads and \SI{11.4}{\mega\byte\per\second} in writes, around 8$\times$ the throughput of SPI.
This approaches the interface bandwidth of \SI{12.5}{\mega\byte\per\second}, accounting for protocol overhead and register access time.
\emph{Bare} shows performance of \SI{6.3}{\mega\byte\per\second} and \SI{9.1}{\mega\byte\per\second} for reads and writes, respectively; 6.5$\times$ and 4.5$\times$ the SPI performance.

Under Linux, we can see the performance uplift gained by skipping fences when accessing \gls{sdhc} registers:
The performance improves from \SI{224}{\kilo\byte\per\second} and \SI{159}{\kilo\byte\per\second} in reads and writes, respectively (\emph{\gls{sdhc}}), to \SI{945}{\kilo\byte\per\second} and \SI{485}{\kilo\byte\per\second} (\emph{Opt}).
An improvement of 24.9$\times$ in reads and 11.3$\times$ in writes over SPI.
The scaled system with \emph{Opt} recovers most of the performance.

We present synthesis \lb{results} for the controller's area and the boot driver's contribution to the bootrom area in IHP's open \SI{130}{\nano\metre} node.
While the current SPI peripheral incurs \SI{86}{kGE}, our \gls{sdhc} uses \SI{24.2}{kGE}, or 3.6$\times$ less area.
The contribution to the boot ROM, particularly important due to its low placement density, also shrinks, from 8 to \SI{4.2}{kGE}. %

\section{Outlook}

To further improve performance, we propose using \glspl{cmo} to reduce unnecessary cache invalidation and to mitigate the impact of Cheshire's complex memory system on register access latency.
HPDcache~\cite{fuguet_hpdcache_2023} supports the \gls{cmo} extension, resulting in lighter-weight fences.
Further, it allows the core to issue multiple outstanding accesses to memory-mapped registers, mitigating access latency and enabling loop-unrolling optimizations, thus benefiting other peripherals as well.
Upgrading to \lb{version 3.0 of the} \gls{sdhci} standard would \lb{enable us to implement and expose} a \gls{sd} DMA to the host, massively increasing throughput.

\printbibliography %

\end{document}